# Performance Analysis of Clustering Algorithms for Gene Expression Data

T.Chandrasekhar, K.Thangavel, E. Elayaraja

**Abstract**— Microarray technology is a process that allows thousands of genes simultaneously monitor to various experimental conditions. It is used to identify the co-expressed genes in specific cells or tissues that are actively used to make proteins, This method is used to analysis the gene expression, an important task in bioinformatics research. Cluster analysis of gene expression data has proved to be a useful tool for identifying co-expressed genes, biologically relevant groupings of genes and samples. In this paper we analysed K-Means with Automatic Generations of Merge Factor for ISODATA- AGMFI, to group the microarray data sets on the basic of ISODATA. AGMFI is to generate initial values for merge and Spilt factor, maximum merge times instead of selecting efficient values as in ISODATA. The initial seeds for each cluster were normally chosen either sequentially or randomly. The quality of the final clusters was found to be influenced by these initial seeds. For the real life problems, the suitable number of clusters cannot be predicted. To overcome the above drawback the current research focused on developing the clustering algorithms without giving the initial number of clusters.

——————— ◆ ———————

## 1 INTRODUCTION

CLUSTERING has been used in a number of applications such as engineering, biology, medicine and data mining. Cluster analysis of gene expression data has proved to be a useful tool for identifying co-expressed genes. DNA microarrays are emerged as the leading technology to measure gene expression levels primarily, because of their high throughput. Results from these experiments are usually presented in the form of a data matrix in which rows represent genes and columns represent conditions [12]. Each entry in the matrix is a measure of the expression level of a particular gene under a specific condition. Analysis of these data sets reveals genes of unknown functions and the discovery of functional relationships between genes [18]. The most popular clustering algorithms in microarray gene expression analysis are Hierarchical clustering [11], K-Means clustering [3], and SOM [8]. Of these K-Means clustering is very simple and fast efficient. The K-Means clustering algorithm which is developed by Mac Queen [6]. The K-Means algorithm is effective in producing clusters for many practical applications. One drawback in the K-Means algorithm is that of a priori fixation of number of clusters [2, 3, 4, 17].

Iterative Self-Organizing Data Analysis Techniques (ISODATA) tries to find the best cluster centres through iterative approach, until some convergence criteria are met. One significant feature of ISODATA over K-Means is that the initial number of clusters may be merged or split, and so the final number of clusters may be different from the number of clusters specified as part of the input. In [10] Karteeka Pavan et al proposed an algorithm AGMFI to initialize merge factor for ISODATA. This paper studies an initialization of centroids proposed in [17] for microarray data to get the best quality of clusters.

This paper is organised as follows. Section 2 presents an overview of Existing works K-Means algorithm, Iterative Self – Organizing Data Analysis Techniques and Automatic Generation of Merge Factor for ISODATA (AGMFI) methods. Section 3 describes the centre initialization algorithm. Section 4 describes performance study of the above methods for UCI data sets. Section 5 describes the conclusion and future work.

## 2 RELATED WORK

### 2.1 K- Means Clustering

The main objective in cluster analysis is to group objects that are similar in one cluster and separate objects that are dissimilar by assigning them to different clusters. One of the most popular clustering methods is K-Means clustering algorithm [3, 9, 12, 17]. It is classifies objects to a pre-defined number of clusters, which is given by the user (assume K clusters). The idea is to choose random cluster centres, one for each cluster. These centres are preferred to be as far as possible from each other. In this algorithm mostly Euclidean distance is used to find distance between data points and centroids [6]. The Euclidean distance between two multi-dimensional data points $X = (x_1, x_2, x_3, ..., x_m)$ and $Y = (y_1, y_2, y_3, ..., y_m)$ is described as follows:

$$D(X, Y) = \sqrt{(x_1 - y_1)^2 + (x_2 - y_2)^2 + \cdots + (x_M - y_M)^2}$$

The K-Means method aims to minimize the sum of squared distances between all points and the cluster centre. This procedure consists of the following steps, as described below

- T.Chandrasekhar is with the Computer Science Department, Periyar University, Salem, Tamilnadu-636 011, India, PH- +91-9942925467. E-mail: ch_ansekh80@rediffmail.com
- K.Thangavel is with the Computer Science Department, Periyar University, Salem, Tamilnadu-636 011, India, E-mail: drktvel@yahoo.com
- E. Elayaraja is with the Computer Science Department, Periyar University, Salem, Tamilnadu-636 011, India, E-mail: elayarajaphd.e@gmail.com

**Algorithm 1:** K-Means clustering algorithm [13]

---

**Require**: D = {d$_1$, d$_2$, d$_3$, ..., d$_n$ } // Set of n data points.
        K - Number of desired clusters
**Ensure**: A set of K clusters.
**Steps:**
1. Arbitrarily choose *k* data points from *D* as initial centroids;
2. **Repeat**
    Assign each point d$_i$ to the cluster which has the closest centroid;
    Calculate the new mean for each cluster;
  **Until** convergence criteria is met.

---

Though the K-Means algorithm is simple, it has some drawbacks of quality of the final clustering, since it highly depends on the arbitrary selection of the initial centroids [1].

### 2.2 Iterative Self-Organizing Data Analysis Techniques

ISODATA algorithms variation is to permit splitting and merging of the resulting clusters. Typically, a cluster is split when its variance is above a pre-specified threshold, and two clusters are merged when the distance between their centroids is below another pre-specified threshold [14]. Using this variant, it is possible to obtain the optimal partition starting from any arbitrary initial partition, provided proper threshold values are specified. The well-known ISODATA algorithm uses more clustering technique of merging and splitting clusters. If ISODATA is given the "ellipse" partitioning shown in Fig.1 as an initial partitioning, it will produce the optimal three-cluster partitioning ISODATA will first merge the clusters {A} and {B,C} into one cluster because the distance between their centroids is small and then split the cluster {D,E,F,G}, which has a large variance, into two clusters {D,E} and {F,G}.

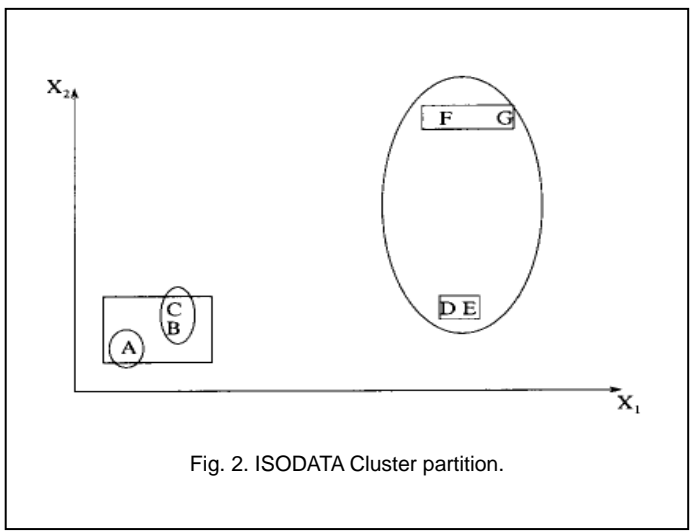

Fig. 2. ISODATA Cluster partition.

**Algorithm 2:** ISODATA algorithm [15]

---

**Input**: D = {d$_1$, d$_2$, d$_3$, ..., d$_n$ } // Set of n data points.
    K - Number of desired clusters.
    $\Theta$n - a threshold value point for discarding cluster.
    $\Theta$s - a threshold value for spilt operation.
    $\Theta$c - a threshold value for merge operations.
**Output**: A set of K clusters.
**Steps**:
1. Select a K- initial partition of the patterns with a fixed number of clusters and cluster centers;
2. Assign each pattern to its closest cluster center and compute the new cluster centers as the centroids of the clusters. Repeat this step until convergence is achieved, i.e., until the cluster membership is stable;
3. Merge and split clusters based on some heuristic information, optionally repeating step 2.

---

### 2.3 Automatic Generation of Merge Factor for ISODATA (AGMFI) Algorithm

The clusters produced in the K-Means clustering are further optimized by ISODATA algorithm. Some of the parameters are fixed by user during the merging and partitioning the clusters. In [10], Automatic Generation of Merge Factor is proposed to initialize merge factor for ISODATA. AGMFI uses different heuristics to determine when to split. Decision of merging is done based upon merge factor which is the function of distances between the clusters. The step by step procedure of AGMFI is given here under.

---

**Algorithm 3:** The AGMFI algorithm [10]

---

**Input**: D = {d$_1$, d$_2$, d$_3$, ..., d$_n$ } // Set of n data points.
    K - Number of desired clusters.
    m- minimum number of samples in a cluster.
    n – maximum number of iterations.
    $\Theta$s – a threshold value for spilt_size.
    $\Theta$c - a threshold value for merge_size.
**Output**: A set of K clusters.
**Steps**:
1. Identify clusters using K-Means algorithms;
2. Find the inter distance in all other cluster to minimum average inter distances clusters point in C;
3. Discard the m and merging operations of cluster ≥ 2*K, If n is even go to step 4 or 5;
4. Distance between two centroids < C, merge the cluster And update centroid, otherwise repeat up to K/2 times;
5. K ≤ K/2 or n is odd go to step 6 or 7;
6. Find the standard division of all clusters that has exceeds S * standard division of D;
7. Executed n times or no changes occurred in clusters since the last time then stop, otherwise take the centroids of the clusters as new seed points and find the clusters using K- Means and go to step 3.

The main difference between AGMFI and ISODATA is ISODATA uses heuristic values to merge the clusters, AGMFI generates automatically and the choice of c is not fixed but is to be decided to have better performance. The distance measure used here is the Euclidean distance. To assess the quality of the clusters, we used the silhouette measure proposed by Rousseeuw [14].

## 3 CLUSTER CENTRE INITIALIZATION ALGORITHM (CCIA)

Performance of iterative clustering algorithms which converges to numerous local minima depends highly on initial cluster centers. Generally initial cluster centers are selected randomly. In this section, the cluster centre initialization algorithm is studied to improve the performance of the K-Means algorithm.

---

**Algorithm 4:** Finding the initial centroids [17]

---

**Inpu**t:  D = {$d_1, d_2, ..., d_n$} // set of *n* data items
        K // Number of desired clusters
**Output**: A set of K initial centroids.

**Steps**:
1. Set m = 1;
2. Compute the distance between each data point and all other data points in the set D;
3. Find the closest pair of data points from the set D and form a data point set $A_m$ ($1 \leq m \leq K$) which contains these two data points, Delete these two data points from the set D;
4. Find the data point in D that is closest to the data point set $A_m$, Add it to $A_m$ and delete it from D;
5. Repeat step 4 until the number of data points in $A_m$ reaches 0.75 * (n/K);
6. If m < K, then m = m+1, find another pair of data points from D between which the distance is the shortest, form another data-point set $A_m$ and delete them from D, Go to step 4;
7. For each data point set $A_m$ ($1 \leq m \leq K$) find the arithmetic mean of the vectors of data points in $A_m$, these means will be the initial centroids.

---

## 4 EXPERIMENTAL ANALYSIS AND DISCUSSION

The following data sets are used to analyse the methods studied in sections 2 and 3.

### 4.1 Serum Data

This data set is described and used in [10]. It can be downloaded from: http://www.sciencemag.org/feature/data/984559.shl and corresponds to the selection of 517 genes whose expression varies in response to serum concentration inhuman fibroblasts.

### 4.2 Yeast data

This data set is downloaded from Gene Expression Omnibus-databases. The Yeast cell cycle dataset contains 2884 genes and 17 conditions. To avoid distortion or biases arising from the presence of missing values in the data matrix we removal all the genes that had any missing value. This step results in a matrix of size 2882 * 17. The proposed matrix contains integer in the range of 1 to 500.

### 4.3 Simulated Data

It is downloaded from http://www.igbmc.u-strasbg.fr/projets/fcm/y3c.txt. The set contains 300 Genes [3]. Above the microarray data set values are all normalized in every gene average values zero and standard deviation equal to 1.

### 4.4 Comparative Analysis

The K-Means, CCIA with K-Means and AGMFI are applied on serum data set when numbers of clusters are taken as 10 and 7 times running to EIAGMFI clusters data into 6.

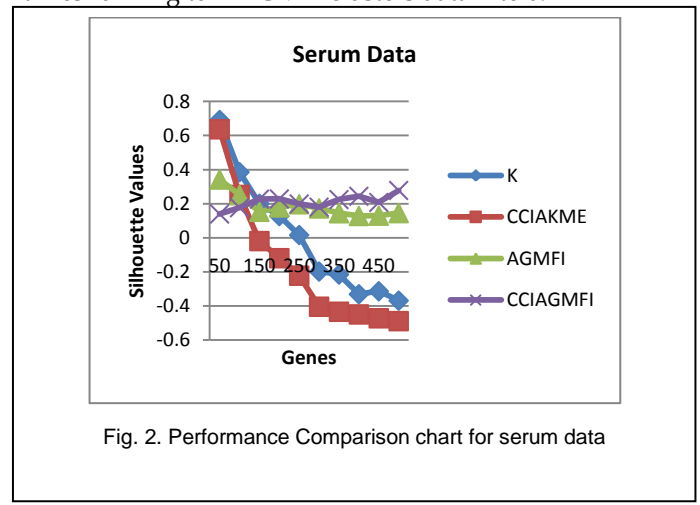

Fig. 2. Performance Comparison chart for serum data

K-Means, CCIA with K-Means and AGMFI are applied on Yeast data set when number of clusters initialized to 10 and 7 times running on EIAGMFI clusters data into 6.

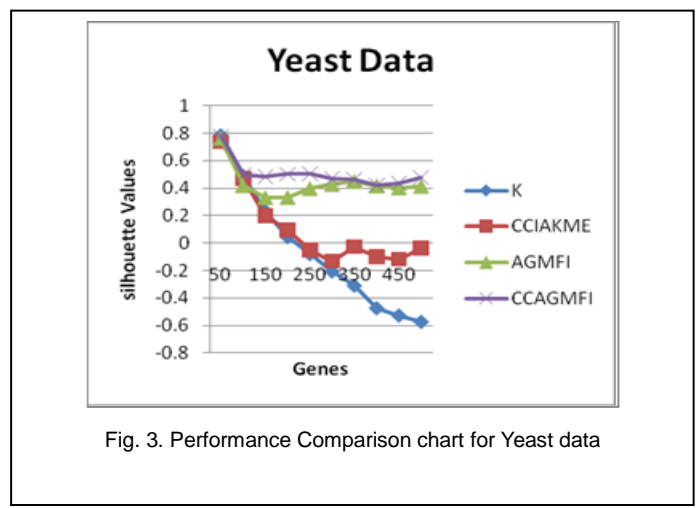

Fig. 3. Performance Comparison chart for Yeast data

The K-Means, CCIA with K-Means and AGMFI are applied on simulated data set when no of clusters initialized to 10 and 7 times running to EIAGMFI clusters data into 5.

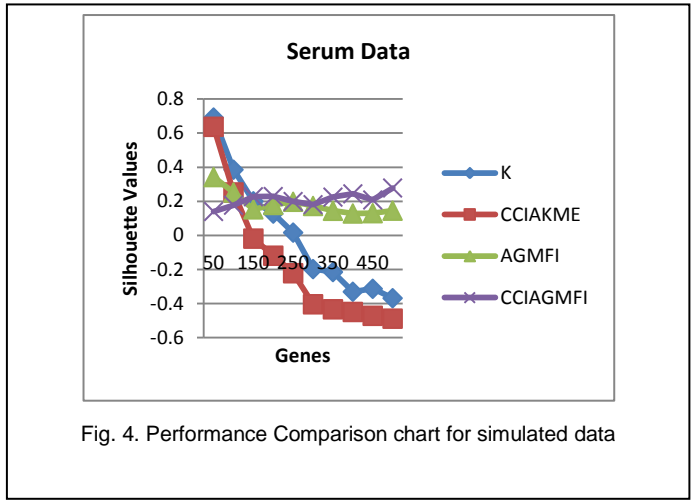

Fig. 4. Performance Comparison chart for simulated data

TABLE 1
COMPARATIVE ANALYSIS OF CLUSTERING QUALITY

| Data set | Initial number of cluster | Finalized number of cluster | Cluster Quality by K-Means | Cluster Quality by CCIA with K-Mean | Cluster Quality by AGMFI | Cluster Quality by EIAGMFI |
|---|---|---|---|---|---|---|
| Serum | 10 | 6 | -0.013 | -17.162 | 18.476 | 21.101 |
| simulated | 10 | 5 | 38.407 | 25.962 | 54.347 | 57.552 |
| Yeast | 10 | 6 | -6.072 | 10.425 | 43.559 | 50.397 |

## 5 CONCLUSION

In this paper AGMFI was studied to improve the quality of clusters. The Evaluation of Improved Automatic Generation of Merge Factor for Clustering Microarray Data based on K-Means and AGMFI clustering algorithms were also studied. One of the demerits of AGMFI is random selection of initial seed point of desired clusters. This was overcome with CCIA for finding the initial centroids algorithms to avoidance for initial values at random. Therefore, the EIAGMFI algorithm not depending upon the any choice of the number of cluster and automatic evaluation initial seed of centroids it produces different better results. Both the algorithms were tested with gene expression data.